\documentclass[preprint,aps]{revtex4}
\usepackage{yfonts}
\usepackage [cp1251]{inputenc}
\makeatletter\AtBeginDocument{\let\@elt\relax}\makeatother
\usepackage{amsmath}
\usepackage{bm}
\usepackage{epsfig}
\usepackage{color}
\usepackage[unicode=true,colorlinks=true,citecolor=blue,urlcolor=blue]{hyperref}

\topmargin - 2.5 cm 
\usepackage{amsmath}
\usepackage{amssymb}

\usepackage{braket}
\graphicspath{{img/}}

\begin{document}
\title{Influence of Coulomb interaction on interband photogalvanic effect in semiconductors}

\author{G.\,V. Budkin,   E.\,L. Ivchenko}
\affiliation{Ioffe Institute, 194021 St.~Petersburg, Russia
}
\author{e-mail: budkin@mail.ioffe.ru}
\begin{abstract}
The ballistic and shift contributions to the interband linear photogalvanic effect are calculated in the same band structure model of a noncentrosymmetric semiconductor. The calculation uses a two-band generalized Dirac effective Hamiltonian with the off-diagonal components containing ${\bm k}$-dependent terms of the first and second order. The developed theory takes into account the Coulomb interaction between the photoexited electron and hole. It is shown that in typical semiconductors the ballistic photocurrent $j^{({\rm bal})}$ significantly exceeds the shift current $j^{({\rm sh})}$: the ratio $j^{({\rm sh})}/j^{({\rm bal})}$ has the order of $a_B/ \ell$, where $a_B$ is the Bohr radius and $\ell$ is the mean free path of photocarriers due to their quasi-momentum scattering.
\vspace{3 mm}

\end{abstract}

\maketitle

\section{Introduction} 
Under the influence of an alternating electromagnetic field, dc photocurrents can arise in macroscopically homogeneous crystals or laterally homogeneous two-dimensional semiconductor structures without a spatial inversion center. Such phenomena are usually called photogalvanic effects. The purpose of this work is to calculate, within the framework of a unified model of the band structure of a semiconductor, the ballistic and shift contributions to the linear photogalvanic effect (LPGE) taking into account the Coulomb interaction between photoexcited electron and hole, and to compare these contributions with each other. The first contribution is due to an asymmetry of the distribution of charge photocarriers in the quasi-momentum space \cite{BelinicherUFN,Alper0,Alperov}, and the second one arises due to the shift of electron wave packets in the real space in optical transitions \cite{Shift,Sturmanbook,Ivchenko}. An important property of the LPGE is that, in direct optical transitions without taking into account an additional scattering of the electron-hole pair, no ballistic contribution arises; for example, it is necessary to take into account the scattering of the electron and hole on each other (Coulomb contribution), on lattice vibrations (phonon contribution), on lattice defects or other charge carriers.

We set ourselves the task of eliminating the existing contradiction in the estimates of the relative roles of the ballistic and shift photocurrents, $j^{({\rm bal})}$ and $j^{({\rm sh})}$, respectively, generated during optical transitions between the valence and conduction bands. In a special methodological note \cite{SturmanUFN} it is pointed out that, under interband transitions, the ballistic contribution to the LPGE is dominant. On the contrary, in the later published works \cite{Dai,RappeReview} it is stated that the ballistic current arising with allowance for the Coulomb interaction of the electron and hole is significantly smaller than the shift contribution. This contradiction, as well as the presence of a large number of works devoted to the calculation of only the shift contribution to the LPGE, see, e.g. \cite{FWang,Tan,Cook,Fregoso,CWang,Wannier,Kim,Fei,Sau,Chan2021,Krishna,Schankler,Tian,
Moire,Zhu}, requires the consideration of the both contributions to the photocurrent for the same fixed model of the semiconductor band structure.

In this paper the electron-hole Coulomb interaction is taken into account while considering both the ballistic and shift photocurrents. The first work on the exact accounting of this interaction in calculation of the current $j^{({\rm bal})}$ was published forty five years ago \cite{Entin1979}. Unlike that work, we take into account that the matrix elements of the velocity operator calculated between the Coulomb functions of the continuous spectrum  $\psi^{(+)}_{{\bm k}'}$ and $\psi^{(+)}_{\bm k}$ \cite{Landau} have off-diagonal components with respect to ${\bm k}'$ and ${\bm k}$. Here, an analytical expression for the shift photocurrent is also derived for the first time with account for the Coulomb interaction. Previously, such an account was reduced only to writing the shift photocurrent as a sum of a general form without transforming it to an analytical formula containing the parameters of the material band structure \cite{Chan2021}.
\section{Dirac~Hamiltonian~in~a~semiconductor~without~an inversion~center}
We use a model of two-band electron structure of a semiconductor with the tetrahedral symmetry T$_d$ with spinor basis functions at the $\Gamma$-point which transform according to the representations $\Gamma_6$ in the conduction band and $\Gamma_7$ in the valence band. These basis functions are expressed in the following form through the Bloch orbital functions $S,X,Y,Z$ and spin columns $\alpha, \beta$ (with the spin projection $+1/2$ and $-1/2$ on the axis $z \parallel [001]$)
\begin{eqnarray}
\label{basise-}
&& \psi_{\Gamma_6, 1/2}  =~ {\rm i} \alpha S\;, \psi^{(e)}_{\Gamma_6, -1/2} = {\rm i} \beta S\:,\\
&& \psi_{\Gamma_7, 1/2} = -\frac{1}{\sqrt{3}} [ \alpha Z + \beta (X+ {\rm i} Y) ]\: , \\
&& \psi_{\Gamma_7, -1/2} =\frac{1}{\sqrt{3}}\: [ \beta Z -
\alpha (X- {\rm i} Y) ] \:. \nonumber \label{basish+}
\end{eqnarray}
In this basis, the generalized Dirac Hamiltonian takes the form \cite{AronovPikus,winkler_book}
\begin{equation} \label{Hk}
\hat{H}_0 = \begin{pmatrix}
 E_g/2&P {\bm \sigma}{\bm k} +{\rm  i} Q \bm{\sigma}\bm{\pi}\\
  P {\bm \sigma}{\bm k} - {\rm i} Q \bm{\sigma}\bm{\pi} &- E_g/2
\end{pmatrix} 
\:,
\end{equation}
where $P$ is a real band parameter $({\rm i} \hbar/\sqrt{3} m_0) \langle S | \hat{p}_x| X  \rangle$, $\bm{\pi}= (k_y k_z, k_x k_z, k_x k_y)$, the Cartesian coordinates $x,y,z$ are directed along the crystallographic axes $[100],  [010], [001]$, respectively, ${\bm \sigma}$ is a three-dimensional pseudovector whose components are the Pauli matrices; off-diagonal terms, quadratic in ${\bm k}$ and determined by the band parameter $Q$, describe the inversion asymmetry, they arise due to the contribution of remote bands in the L\"owdin procedure \cite{Lowdin}, which allows reducing the multi-zone Hamiltonian to a matrix (\ref{Hk}) of the $4\times 4$ dimension.

We consider the optical transitions near the forbidden band $E_g$ and assume the light frequency $\omega$ to satisfy the inequality
\begin{equation} \label{frequency}
\hbar \omega - E_g \ll E_g\:.
\end{equation} 
In this case, it is sufficient to limit ourselves to the parabolic dispersion of electrons in the conduction band $c$ and the valence band $v$:
\begin{equation} \label{energies}
\varepsilon_{c{\bm k}} = - \varepsilon_{v{\bm k}} \equiv \varepsilon_{h {\bm k}} = \frac{E_g}{2} + \frac{\hbar^2 k^2}{2 m^*}\:,
\end{equation}
where the single-particle effective mass $m^* = \hbar^2 E_g/ 2 P^2$, and the electron-hole reduced mass $\mu = m^*/2$. Note, however, that the photocurrent considered here is proportional to the band asymmetry parameter $Q$, which will be taken into account in the matrix elements of the interband transitions.
\subsection{Continuum electron-hole states}
In the effective mass method, the two-particle wave function of an electron and a hole with zero quasi-momentum of the centre of mass can be represented in general form as
\begin{equation} \label{Psiss}
\Psi_{s_e,s_h} = \psi({\bm r}) u_{\Gamma_6,s_e}({\bm r}_e) u^{(h)}_{\Gamma_7,s_h}({\bm r}_h)\:.
\end{equation}
Here $\psi({\bm r})$ is a smooth envelope function of the difference variable ${\bm r} = {\bm r}_e - {\bm r}_h$, $u_{\Gamma_6,s_e}({\bm r}_e)$ and $u^{(h)}_{\Gamma_7,s_h}({\bm r}_h) = - K u_{\Gamma_7, - s_h}({\bm r}_h)$ are the electron and hole Bloch periodic amplitudes at the extremum point ${\bm k} = 0$, $s_e$ and $s_h$ are the spin projections $\pm 1/2$ onto the $z$-axis direction, $K$ is the time inversion operator that relates the states in the electron and hole representations: $K = - {\rm i} \sigma_y K_0$, $\sigma_y$ being the Pauli matrix, and $K_0$ being the complex conjugation operator. Since the energies (\ref{energies}) do not depend on the spin states, the smooth envelope is also independent of the indices $s_e,s_h$. For convenience, we will set the crystal normalization volume $V$ equal to unity.

As eigenstates of the Coulomb problem $\Psi_{s_e,s_h, {\bm k}} \equiv |s_e,s_h, {\bm k} \rangle$, we choose the smooth envelopes $\psi^{(+)}_{\bm k}(\bm r)$, which at large distances converges to the plane waves ${\rm exp}({\rm i}{\bm k} {\bm r})$. Their expansion in spherical waves has the form \cite{Landau}
\begin{equation} \label{exponentC}
\psi^{(+)}_{\bm k}(\bm r)= \frac{2 \pi}{k} \sum\limits_{l = 0}^{\infty} \sum\limits_{m = - l}^{l}  
{\rm i}^l {\rm e}^{{\rm i} \delta_l} R_{kl}(r)Y^*_{l,m}\left( \frac{\bm k}{k} \right)Y_{l,m}\left( \frac{\bm r}{r} \right) \:,
\end{equation}
where the phase $\delta_l = {\rm arg}\hspace{1 mm} \Gamma [l+ 1 - ({\rm i}/ka_B)]$, and the radial function $R_{kl}(r)$ has the unit of the inverse length and, when choosing the normalization according to \cite{Landau}, is equal to
\begin{equation}  \label{Rkla}
R_{kl}(r) = \frac{C_{kl}}{(2l + 1)!} (2kr)^l{\rm e}^{- {\rm i} k r} F \left( l+ 1 + \frac{\rm i}{ka_B}, 2l + 2, 2 {\rm i} kr \right)\:.
\end{equation}
Here $F(\alpha, \beta, z)$ is the degenerate hypergeometric function, the exciton Bohr radius $a_B = \kappa \hbar^2/e^2 \mu$ is introduced with $\kappa$ being the permittivity of the medium. Let us present expressions for the coefficients $C_{kl}$ with the orbital moment $l = 0$ and 1
\begin{equation}
C_{k0} = 2k \sqrt{\cal Z}\:,\: C_{k1} = \sqrt{1 + \frac{1}{(ka_B)^2}}\ C_{k0}\:,
\end{equation}
where the Sommerfeld factor equals to
\begin{equation} \label{Sommer}
{\cal Z} = \frac{X}{1 - {\rm exp}(-X)}\:,\: X = \frac{2 \pi}{k a_B}\:.
\end{equation}
Other normalization coefficients can be found in the literature
\begin{equation} \label{relation}
R^{LL}_{kl}(r) = \sqrt{2 \pi} R^G_{kl}(r) = \sqrt{2 \pi} k R^{VP}_{El}(r) = \sqrt{ \frac{2}{\pi k}} R^{K}_{kl}(r) = 2k R^{Ell}(r)\:.
\end{equation}
Here the function $R_{kl}(r)$ normalized according Landau and Lifshitz and entering Eq. (\ref{exponentC}) is denoted for clarity as $R^{LL}_{kl}(r)$, and the remaining functions with superscripts $G, VP, K$ and $Ell$ are introduced in the articles \cite{Gordon,Veniard, Komninos2012, Elliott}, respectively. When using radial functions with a different normalization in the expansion (\ref{exponentC}), one must multiply this expansion by the corresponding coefficient in the relations (\ref{relation}). The excitation energy of the electron-hole state (\ref{exponentC}) has a parabolic dispersion
\begin{equation} \label{pair-energy}
E_{\bm k} = E_g + \frac{\hbar^2 k^2}{2 \mu}\:.
\end{equation}

In what follows, we will also use the expansion of the Coulomb wave function in terms of the states of non-interacting electron and hole,
\begin{equation} \label{exp2}
| s_e,  s_h, {\bm k} \rangle = \Psi^{(+)}_{s_e,s_h,{\bm k}} = \sum\limits_{\bm q} C^{({\bm k})}_{\bm q}  | s_e, {\bm q}; s_h, -{\bm q}; {\rm free} \rangle\:,
\end{equation}
where
\[
| s_e, {\bm q}; s_h, -{\bm q}; {\rm free} \rangle = {\rm e}^{{\rm i} {\bm q}{\bm r}}u_{\Gamma_6,s_e}({\bm r}_e) u^{(h)}_{\Gamma_7,s_h}({\bm r}_h)\:,
\]
$C^{({\bm k})}_{\bm q}$ is a Fourier transform of the envelope $\psi^{(+)}_{\bm k}({\bm r})$.
\section{Two contributions to the linear photogalvanic current}
The photoinduced electric current is contributed by the photon drag effect, circular and linear photogalvanic effects (PGE). The first contribution arises due to the transfer of photon momentum to free charge carriers, it is proportional to the wave vector of light. The second contribution is due to the transformation of the angular momentum of circularly polarized photons into the translational motion of free electrons or holes and is proportional to the degree of circular polarization of the radiation $P_{\rm circ}$. The linear PGE arises in piezoelectrics, it does not depend on the wave vector of light or the degree of polarization $P_{\rm circ}$ and is usually studied with linear polarization of the exciting light.

In turn, the linear photocurrent consists of the ballistic and shift contributions
\[
{\bm j} = {\bm j}^{({\rm bal})} + {\bm j}^{({\rm sh})}\:.
\]
Without taking into account the Coulomb interaction, these currents are calculated in the single-particle approximation using the formulas
\begin{eqnarray} \label{rho} &&{\bm j}^{({\rm bal})} = e \sum_l \sum_{{\bm k}s's} {\bm v}_{ls,ls'} ({\bm k}) \overline{\rho}_{ls',ls} ({\bm k})\:,\\
&& {\bm j}^{({\rm sh})} = e \sum_{l \neq l'} \sum_{{\bm k}s' s} {\bm v}_{ls,l's'} ({\bm k}) \overline{\rho}_{l's',ls} ({\bm k})\:,  \nonumber
\end{eqnarray}
where $l',l$ are the indices of the $c$ and $v$ bands, $s,s'$ are the spin indices, ${\bm v}_{ls,l's'}$  are the matrix elements of the velocity operator, $\overline{\rho}_{l's',ls} ({\bm k})$ is the single-particle density matrix averaged over time. Taking into account the Coulomb interaction, Eqs.  (\ref{rho}) take the form
\begin{subequations}
\begin{align}
\label{bal}
& {\bm j}^{({\rm bal})} = e  \sum_{{\bm k}{\bm k'} s_e s_h} {\bm v}_{{\bm k}' {\bm k}} \overline{\rho}_{s_e, s_h, {\bm k}; s_e, s_h, {\bm k}'} + {\rm c.c.}\:, \\
&{\bm j}^{({\rm sh})} = e \sum_{{\bm k} s_e s_h} \overline{\left\langle 0 \left| \hat{\bm v} \right|s_e, s_h, {\bm k} \right\rangle \rho_{s_e, s_h, {\bm k};0}} + {\rm c.c.} \:, \label{sh}
\end{align}
\end{subequations}
where
\begin{equation} \label{k'vk}
{\bm v}_{{\bm k}' {\bm k}}  = \int \psi^{(+) *}_{{\bm k}'}({\bm r}) \left(- {\rm i} \frac{\hbar}{\mu} \frac{\partial}{\partial {\bm r}}\right) \psi^{(+) }_{\bm k}({\bm r}) d {\bm r}\:,
\end{equation}
$|0\rangle$ is the ground state of the crystal (the filled valence band and empty conduction band).
A brief derivation of Eqs. (\ref{bal}), (\ref{k'vk}) is given in Appendix \ref{Append1}. An expression for the matrix element of the operator $\hat{\bm v}$ in (\ref{sh}) in terms of the coefficients $C^{({\bm k})}_{\bm q}$ is also given there.

In a bulk semiconductor of the T$_d$ symmetry, the linear photogalvanic effect, both ballistic and shift, is phenomenologically described by \cite{Entin}
\begin{equation}
j_i = \chi e_{i+1} e_{i+2} {\cal E}_0^2\:.
\end{equation}
Here ${\cal E}_0$ is the real amplitude of the electric field of the radiation, $\bm{e}$ is the unit vector of linear polarization, $i = x,y,z$, and a cyclic permutation of coordinates is assumed, $x \to y \to z \to x$. For definiteness, we will consider the polarization 
\begin{equation} \label{polariz}
{\bm e}= \frac{1}{\sqrt{2}} (1,1,0) \:, 
\end{equation}
for which a photocurrent is induced in the $z$ direction. Neglecting the wave vector of the photon, the vector potential and the electric field oscillate in time according to
\begin{equation} \label{field}
\bm{\mathcal A} (t) =   {\bm e} {\cal A}_0  \left( {\rm e}^{- {\rm i} \omega t} + {\rm e}^{{\rm i} \omega t} \right)\:,\:\bm{\mathcal E} (t) = 2 {\bm e} {\cal E}_0  \sin{\omega t}\:,\: {\cal A}_0 = \frac{c}{\omega} {\cal E}_0\:.
\end{equation}
In this case, the operator of interaction between light and electrons takes the form
\[
\hat{V}(t) = \hat{V} ( {\rm e}^{- {\rm i} \omega t} + {\rm e}^{{\rm i} \omega t} )\:,
\]
where
\begin{equation} \label{V+}
\hat{V}= - \frac{e}{\omega} (\hat{\bm v}_0 \cdot {\bm e}) {\cal E}_0 \:,\: \hat{\bm v}_0 = \frac{1}{\hbar} \frac{\partial H_0}{\partial {\bm k}}\:. 
\end{equation}

\subsection{The optical excitation matrix elements}
In the absence of Coulomb interaction, under the condition (\ref{frequency}) and in the polarization (\ref{polariz}), we have for the matrix elements of the optical transitions in the electronic representation 
\begin{align}
\label{me_bulk}
V_{c,\pm \frac{1}{2}, \bm{k}; v, \pm, \frac{1}{2}, \bm{k}}&=\mp {\rm i} \dfrac{e {\cal E}_0}{\sqrt{2} \hbar \omega}  Q(k_x+k_y)\:,\nonumber\\
V_{c, \pm, \frac{1}{2}, \bm{k}; v, \mp \frac{1}{2},\bm{k}}&= - \frac{1 \mp {\rm i}}{\sqrt{2}} \dfrac{e {\cal E}_0}{ \hbar\omega} (P+ {\rm i} Q k_z)\:.
\end{align}
Only the transitions $(v, -1/2, {\bm k}) \to (c, 1/2, {\bm k})$ and $(v, 1/2, {\bm k}) \to (c, -1/2, {\bm k})$, whose matrix elements contain both coefficients $P$ and $Q$, lead to a photocurrent.
Note that, while deriving Eqs. (\ref{me_bulk}), we took into account the condition (\ref{frequency}), under which the ${\bm k}{\bm p}$-mixing of the conduction and valence band states  can be neglected, the single-particle initial and final states have the form
\begin{equation} \label{ucuv}
\psi_{c,s,{\bm k}}({\bm r})={\rm e}^{{\rm i} {\bm k}{\bm r} } u_{\Gamma_6,s}({\bm r})\:,\: \psi_{v, s,{\bm k}}({\bm r})= {\rm e}^{{\rm i} {\bm k}{\bm r} } u_{\Gamma_7, s}({\bm r})
\end{equation}
and the matrix elements (\ref{me_bulk}) do not contain terms of the second or higher order in ${\bm k}$.

Applying the Elliott theory \cite{Elliott}, we can generalize Eq. (\ref{me_bulk}) to transitions from the ground state $|0\rangle$ to the electron-hole Coulomb state $| s_e= \pm 1/2, s_h = \pm 1/2, {\bm k} \rangle$
\begin{equation} \label{MEs}
\langle \pm 1/2, \pm 1/2, {\bm k} | \hat{V} | 0 \rangle \equiv  V_{\pm \frac12, \pm \frac12, {\bm k}; 0}  = -  \frac{1 \mp {\rm i}}{\sqrt{2}} \dfrac{e {\cal E}_0}{ \hbar\omega} ({\rm e}^{-{\rm i} \delta_0} P + {\rm i} {\rm e}^{-{\rm i} \delta_1}Q k_z) \sqrt{\cal Z}\:,
\end{equation} 
where  
\[
{\cal S} = \sqrt{1 + \frac{1}{(k a_B)^2}}
\]
and the Sommerfeld factor is defined according to Eq.~(\ref{Sommer}). For large values of $ka_B$, the factors ${\cal Z}$ and ${\cal S}$  tend to unity, the phases $\delta_0$ and $\delta_1$ tend  to zero, and Eq.~(\ref{MEs}) converges to Eq. (\ref{me_bulk}) (with account for the different signs of the spin projection in the hole and electron representations).

It follows from Eq. (\ref{MEs}) that the main contribution to the light absorption probability per unit time per unit volume is equal to
\begin{equation} \label{W}
W = \frac{4\pi}{\hbar} \left( \frac{eP}{\hbar \omega}\right)^2  g(\hbar \omega) {\cal Z} {\cal E}_0^2\:,
\end{equation} 
where the reduced density of states is given by
\begin{equation} \label{gE}
g(\hbar \omega) = \sum_{\bm k} \delta(\hbar \omega - E_{\bm k}) = \frac{\mu k (\omega)}{2 \pi^2 \hbar^2}\:,\:  k (\omega) = \sqrt{\frac{2 \mu(\hbar \omega - E_g)}{\hbar^2}}\:. 
\end{equation} 
\section{Ballistic photocurrent}
To calculate the ballistic current, we need to find the density matrix $\overline{\rho}_{s_e s_h {\bm k}; s_e s_h {\bm k}'}$ and the matrix element of the velocity operator ${\bm v}_{{\bm k}' {\bm k}}$. We will do this successively.
\subsection{The two-particle  density matrix} 
For brevity, we denote the ground state of the crystal as $| 0 \rangle$ and the excited states as $| s_e,s_h, {\bm k} \rangle$ by one index $n,n'$ or $m$.
The density matrix $\rho_{nn'}(t) = \rho^*_{n'n}(t)$ satisfies the set of equations
\begin{equation} \label{rhot}
    \left[\varepsilon_{n'} - \varepsilon_n+{\rm i} \hbar (\gamma_n + \gamma_{n'})  + {\rm i}\hbar{\partial \over \partial t} \right]\rho_{nn'}(t) = \sum\limits_m [V_{nm}(t) \rho_{mn'}(t) - \rho_{nm}(t)V_{mn'}(t)]\:,
\end{equation}
where $\varepsilon_n$ is the energy of the electron system in the state $n$, $V_{mn}(t)$ is the matrix element of the operator of interaction with the electromagnetic field. For the ground state $| 0 \rangle$ the damping $\gamma_0 = 0$, for the excited states the parameter $\gamma_n \equiv \gamma$ takes into account the scattering of the electron-hole pair on impurities or phonons. For the linearly polarized light (\ref{field}) we have
\[
V_{mn}(t) = V_{mn} \left( {\rm e}^{ - {\rm i} \omega t } + {\rm e}^{ {\rm i} \omega t} \right)\:.
\]
In an intrinsic semiconductor at low temperature the initial density matrix has one non-zero component
\begin{equation} \label{rho0}
\rho^{(0)}_{nn'} = \delta_{n 0} \delta_{n'0} \:.
\end{equation}

In the first order of perturbation theory, the time dependence of the density matrix has the form
\begin{equation}
\rho^{(1)}_{n 0} (t)= \rho^{(1)*}_{0n} (t) = \rho^{(+1)}_{n0} {\rm e}^{- {\rm i} \omega t} + \rho^{(-1)}_{n0} {\rm e}^{{\rm i} \omega t}\:, 
\end{equation}
where $n$ is any excited state. Substituting this expression for the density matrix into the left-hand side of Eq. (\ref{rhot}), and the expression (\ref{rho0}) into the right-hand side, we find
\begin{eqnarray} \label{first}
&&\rho^{(+1)}_{n0} = \frac{V_{n0} }{ \hbar \omega - E_n + {\rm i} \hbar \gamma}\:,\\ &&\rho^{(-1)}_{0n} = \frac{V_{0n}}{\hbar \omega - E_n - {\rm i} \hbar \gamma}  \:, \nonumber
\end{eqnarray}
where it is taken into account that the difference $\varepsilon_n - \varepsilon_0$ is the excitation energy $E_n$, defined according to (\ref{pair-energy}). The non-resonant terms $\rho^{(-1)}_{n0}$ and $\rho^{(+1)}_{0n}$ are not presented since they make no contribution to the ballistic photocurrent.

For the second order of perturbation theory, after averaging over time, we obtain for the components of the density matrix with $n,n'\neq 0$
\begin{equation} \label{rho2nn'}
\overline{\rho}^{(2)}_{nn'} =  \frac{V_{n0} V_{0n'}}{E_{n'} - E_n + 2 {\rm i} \hbar \gamma} \left( \frac{1}{\hbar \omega - E_{n'} - {\rm i} \hbar \gamma} - \frac{1}{\hbar \omega - E_n + {\rm i} \hbar \gamma} \right)\:.
\end{equation}
Replacing $n$ with $s_e,s_h,{\bm k}$, $n'$ with $s_e,s_h,{\bm k}'$ and the energy denominators with delta-functions, we finally find
\begin{equation} \label{rho2a}
\overline{\rho}^{(2)}_{s_e,s_h,{\bm k};s_e,s_h, {\bm k}'} = {\rm i} \pi \frac{ V_{s_e,s_h,{\bm k};0} V_{0; s_e,s_h,{\bm k}'} }{
E_{{\bm k}'} - E_{\bm k} + 2 {\rm i} \hbar \gamma } \left[\delta( \hbar \omega - E_{{\bm k}'} ) +\delta(\hbar \omega - E_{\bm k}) \right]\:.
\end{equation}
The contribution to the photocurrent is made by the odd part of the product
\begin{equation} \label{asymm}
\left( V_{\pm \frac12,\pm \frac12,{\bm k};0} V_{0; \pm \frac12,\pm \frac12,{\bm k}'} \right)_{\rm odd} = {\rm i} \left( \frac{e {\cal E}_0}{\hbar \omega} \right)^2 PQ  \left( {\rm e}^{ {\rm i} (\delta'_0 - \delta_1)}{\cal S} k_z - {\rm e}^{-{\rm i} (\delta_0 - \delta'_1)}{\cal S}' k'_z \right) \sqrt{{\cal Z}{\cal Z}'}\:,
\end{equation}
where $\delta'_l = \delta_l(k')$, ${\cal Z}'= {\cal Z}(k')$, ${\cal S}'={\cal S}(k')$. It follows then that the summing over spins in Eq. (\ref{bal}) can be replaced by doubling the right-hand side of Eq.  (\ref{asymm}).
\subsection{Matrix element of the velocity operator}
Using the relationship between the velocity and coordinate matrix elements, the integral (\ref{k'vk}) can be rewritten as
\begin{equation} \label{k'rk}
\int \psi^{(+) *}_{{\bm k}'}({\bm r}) \left(- {\rm i} \frac{\hbar}{\mu} \frac{\partial}{\partial {\bm r}}\right) \psi^{(+) }_{\bm k}({\bm r}) d {\bm r} = {\rm i} \frac{E_{{\bm k}'} - E_{\bm k}}{\hbar}
\int \psi^{(+) *}_{{\bm k}'}({\bm r}) {\bm r} \psi^{(+) }_{\bm k}({\bm r}) d {\bm r}\:.
\end{equation}
Next, we substitute the expansions (\ref{exponentC}) of the functions $\psi^{(+)}_{\bm k}$ and $\psi^{(+)}_{{\bm k}'}$ into these integrals and take into account that, after integration, only contributions with $l-l' = \pm 1$ will remain non-zero. According to Eq. (\ref{asymm}), the angular dependence of the density matrix (\ref{rho2a}) comes from the factors $k_z$ and $k'_z$. Therefore, in the integrals of Eq. (\ref{k'rk}), the terms with $l=0, l'=1, m'=0$ and $l=1, m=0, l'=0$ must be left only. As a result, we obtain for this part of the matrix element of $z$-coordinate
\begin{equation} \label{Ik'k}
z_{{\bm k}' {\bm k}} \to {\rm i} \frac{\pi}{kk'} \left(   \frac{k_z}{k} {\rm e}^{- {\rm i} (\delta'_0 - \delta_1)} I_{k1,k'0} -  \frac{k'_z}{k'} {\rm e}^{{\rm i} (\delta_0 - \delta'_1)}I_{k0,k'1}\right)\:,
\end{equation}
where
\begin{equation} \label{r10}
 I_{kl,k'l'} = \int\limits_0^{\infty}  R_{kl}(r) R_{k'l'}(r) r^3 dr \:.
\end{equation}

Averaging the product of the expressions (\ref{asymm}) and (\ref{Ik'k}) over directions of the vectors ${\bm k}$ and ${\bm k}'$ we find
\begin{eqnarray} \label{averprod}
&&\int \frac{d \Omega_{\bm k} d \Omega{{\bm k}'}}{(4 \pi)^2}z_{{\bm k}' {\bm k}} \left( V_{\pm \frac12,\pm \frac12,{\bm k};0} V_{0; \pm \frac12,\pm \frac12,{\bm k}'} \right)_{\rm odd} \\ &&  = - \frac{\pi}{3} \left( \frac{e {\cal E}_0}{\hbar \omega} \right)^2 PQ \sqrt{{\cal Z}{\cal Z}'}\left( \frac{{\cal S} I_{k1,k'0}}{k'}  + \frac{{\cal S}'I_{k0,k'1}}{k}\right) \: .\nonumber
\end{eqnarray}
It follows then that after such averaging the phases $\delta_l$ in Eqs. (\ref{asymm}) and (\ref{Ik'k}) cansel  each other and will not arise in the further calculation of the current (\ref{bal}).
\subsection{Calculation of the ballistic photocurrent}
Let us transform the energy denominator in Eq. (\ref{rho2nn'}) to
\begin{equation} \label{denom}
\frac{1}{E_{{\bm k}'} - E_{\bm k} + 2 {\rm i} \hbar \gamma } = \frac{E_{{\bm k}'} - E_{\bm k} - 2 {\rm i} \hbar \gamma }{(E_{{\bm k}'} - E_{\bm k})^2 + (2 \hbar \gamma)^2 }\:.
\end{equation}
Since the integral (\ref{averprod}) is real, the imaginary part of expression (\ref{denom}) does not contribute to the current, and this expression can be replaced by
\[
\frac{E_{{\bm k}'} - E_{\bm k} }{(E_{{\bm k}'} - E_{\bm k})^2 + (\hbar/\tau)^2 }\:,
\]
where $\tau = (2\gamma)^{-1}$ is the scattering time. The sum (\ref{bal}) for the current $j_z$, averaged over the angles of the wave vectors, takes the form
\begin{eqnarray} \label{jzPQ}
&&j^{({\rm bal})}_z = \frac{\pi e}{\hbar} \left( \frac{e E}{\hbar \omega} \right)^2 2PQ \sum_{{\bm k} {\bm k}' } \frac{(E_{{\bm k}'} - E_{\bm k})^2 }{(E_{{\bm k}'} - E_{\bm k})^2 + (\hbar/\tau)^2 }
\frac{\pi}{3} \left( \frac{{\cal S}I_{k1,k'0}}{k'} + \frac{{\cal S}'I_{k0,k'1}}{k}\right)
 \\ && \hspace{3 cm}\times \sqrt{{\cal Z}{\cal Z}'} \left[\delta( \hbar \omega - E_{\bm k} ) +\delta(\hbar \omega - E_{{\bm k}'}) \right]\:. \nonumber
\end{eqnarray}

Thus, to find the ballistic photocurrent, it is necessary to calculate the integrals $I_{k1,k'0}, I_{k0,k'1}$. This calculation is performed in Appendix \ref{Append2}. Taking into account Eqs. (\ref{W}), (\ref{ratiomu}), (\ref{delta}), (\ref{fkk}) we obtain instead of (\ref{jzPQ}) 
\begin{eqnarray} \label{jzf}
j^{({\rm bal})}_z &=& e W  \frac{Q}{P} \frac{2 \pi}{3}\frac{\pi \tau}{\hbar} \left( \frac{\hbar^2}{2 \mu} \right)^2 4k^2 \frac{\mu k}{2 \pi^2 \hbar^2} \frac{2 \pi}{k}\frac{1}{  \pi} \frac{1}{ k a_B}\\ &=& e \frac{Q}{P} W \frac{2}{3}\frac{\tau}{\hbar} \frac{\hbar^2 k}{\mu a_B}  \:. \nonumber
\end{eqnarray}
which is the main result of this work. 

\subsection{Another method to calculate ballistic current}
In this subsection we ignore the influence of the Coulomb interaction on the matrix element of the velocity operator and take this interaction into consideration only in the matrix element of the optical excitation (\ref{MEs}). With this approach, the equation for the photocurrent takes the form
\begin{equation} \label{20}
j_z = \frac{2 \pi}{\hbar} e \tau \sum_{s_e, s_h, {\bm k}}  \frac{\hbar k_z}{\mu} \left\vert V_{s_e, s_h, {\bm k};0} \right\vert^2 \delta (E_k - \hbar \omega) \:.
\end{equation}
Substituting the expressions (\ref{MEs}) into this formula and averaging over the direction of the ${\bm k}$ vector, we obtain
\begin{equation} \label{2}
j_z = \frac{2 \pi}{\hbar} 2 PQ e \tau\sum_{\bm k}  \frac{\hbar k^2}{3 \mu} 2 \sin{(\delta_1 - 
\delta_0)}{\cal ZS} \delta (E_k - \hbar \omega) = \frac23 \frac{Q}{P} W \tau {\cal S} \frac{\hbar k^2}{\mu}  \sin{(\delta_1 -  \delta_0)}\:.
\end{equation}
Considering further that
\[
\sin{(\delta_1 - \delta_0)} = \frac{1}{ka_B {\cal S}}\:,
\]
we arrive at the same formula (\ref{jzf}). Thus, both approaches give the same result.
\section{Shift photocurrent} 
To derive the shift current in the multi-band model \cite{Shift}, it is necessary to substitute the second order of the density matrix $\overline{\rho}^{(2)}_{s_e, s_h, {\bm k};0}$ into (\ref{sh}). Importantly, in the two-band model (\ref{Hk}) with off-diagonal terms non-linear in ${\bm k}$, the velocity operator contains a contribution linear in the electric field
\begin{equation} \label{vHr}
\hat{\bm{v}}=\hat{\bm{v}}_{0}+\delta\hat{\bm{v}}( e^{- {\rm i} \omega t} + e^{{\rm i} \omega t} )\:.
\end{equation} 
Therefore, the expression for the shift current contains an additional contribution from the first-order density matrix and has the form
\begin{equation}  \label{jsh1}
{\bm j}^{({\rm sh})} = e \sum_n \left( \left\langle 0 \left| \hat{\bm v}_0 \right|n \right\rangle \overline{\rho}^{(2)}_{n;0} + \langle 0 | \delta\hat{\bm{v}}  | n \rangle \overline{\rho}^{(+1)}_{n;0}\right)  + {\rm c.c.}
\end{equation}  

Let us start transforming this sum from the second term. The first-order density matrix $\overline{\rho}^{(+1)}_{n;0}$ is defined according to Eq.~(\ref{first}). We expand the factor $ \langle 0 | \delta\hat{\bm{v}} | n \rangle$ in terms of the matrix elements for the free electron-hole pairs
\begin{eqnarray} \label{x}
&& \langle 0 | \delta\hat{\bm{v}}  | s_e, s_h, {\bm k} \rangle = \sum_{\bm q} C^{({\bm k})}_{\bm q } \langle 0 | \delta \hat{\bm{v}}  | s_e, {\bm q}; s_h, - {\bm q}; {\rm free}  \rangle\:. \end{eqnarray}
The identity
\[
\delta {\bm v} = \frac{\rm i}{\hbar} [\hat{V},{\bm r}]
\]
allows us to rewrite the matrix element in the sum in Eq. (\ref{x}) as
\begin{eqnarray} \label{0dv}
&& \langle 0 | \delta \hat{\bm{v}}  | s_e, {\bm q}; s_h, - {\bm q} ; {\rm free} \rangle = \langle c,s_e, {\bm q} | \delta \hat{\bm{v}}  | v, -s_h, {\bm q} \rangle  \\ && = \frac{\rm i}{\hbar} \sum\limits_{l,s,{\bm q}'} (V_{c,s_e, {\bm q}; l,s,{\bm q}} \bm{r}_{l,s , \bm{q}; v, -s_h, \bm{q}'}- \bm{r}_{c,s_e, {\bm q}; l,s, {\bm q}' } V_{l,s, {\bm q}'; -s_h, {\bm q}'} )\:. \nonumber
\end{eqnarray}
The matrix elements of the coordinate are calculated using the formulas \cite{Shift}
\begin{eqnarray}
\label{r_operator_def}
&&\bm{r}_{l,s \bm{q}; l, s', \bm{q}'} = {\rm i} \delta_{s's} \frac{ \partial \delta_{\bm{q}', \bm{q}}}{\partial \bm{q}} ~~~(l=c,v)\:, \\
&&\bm{r}_{c,s \bm{q}; v, s' \bm{q}'}= - \delta_{\bm{q}', \bm{q}} {\rm i} \hbar \frac{ {\bm v}_{ c,s, {\bm q}; v, s', {\bm q}}}{ \varepsilon_{ c, {\bm q}}  - \varepsilon_{v, {\bm q} }} \:. \nonumber 
\end{eqnarray}
Substituting these formulas into the sum in Eq. (\ref{0dv}), we obtain
\begin{equation} \label{ehk0}
\langle 0 | \delta \hat{\bm{v}}  | s_e, {\bm q}; s_h, - {\bm q} ; {\rm free} \rangle = \frac{e}{\hbar} \left[
\frac{ \partial V_{ c, s_e, {\bm q}; v, - s_h, {\bm q} } }{\partial {\bm q}}    + 
\frac{\hbar {\bm v}_{ c,s_e, {\bm q}; v,- s_h, {\bm q}}}{ \varepsilon_{c {\bm q}} - \varepsilon_{v {\bm q}} }  \left( V_{ c, {\bm q}; c, {\bm q} }  - V_{ v, {\bm q}; v, {\bm q}  } \right) \right]^*\:, 
\end{equation}
where
\[
V_{ c, {\bm q}; c, {\bm q} }  - V_{ v, {\bm q}; v, {\bm q}  }  = - \frac{\hbar ({\bm q}{\bm e})}{\mu} \frac{e {\cal E}_0}{\omega}\:.
\]
When replacing the basis functions (\ref{ucuv})
\[
\psi_{c,s,{\bm k}} \to {\rm e}^{{\bm i} \varphi (c,s, {\bm k})} \psi_{c,s,{\bm k}}\:,\: \psi_{v,s,{\bm k}} \to {\rm e}^{{\bm i} \varphi (v,s_e, {\bm k})} \psi_{v,s,{\bm k}}\:,
\]
where $\varphi (l,s, {\bm k})$ is a smooth function of ${\bm k}$, an additional term 
\[
[{\bm \Omega}_{c,s_e}({\bm q}) - \bm{\Omega}_{v,-s_h}({\bm q})]V_{ c, s_e, {\bm q}; v, - s_h, {\bm q} } = {\rm i} \left(\frac{\partial \varphi (c,s_e, {\bm q})}{\partial {\bm q}} - \frac{\partial \varphi (v,-s_h, {\bm q})}{\partial {\bm q}}  \right) V_{ c, s_e, {\bm q}; v, - s_h, {\bm q} }\:,
\]
will appear in the square brackets of Eq.~(\ref{ehk0}) so that the matrix element (\ref{ehk0}) will remain invariant to such a replacement.

Solving Eq. (\ref{rhot}) in the second order, we find
\begin{equation} \label{rho2eh}
\langle n  | \overline{\hat{\rho}}^{(2)} | 0 \rangle = - \frac{\langle n |\overline{ [\hat{V},\hat{\rho}^{(1)}]} | 0 \rangle}{E_n} = - \frac{1}{E_n} \sum_{m}  V_{nm}  \bar{\rho}^{(+1)}_{m0} \:.
\end{equation}

Now we represent the current (\ref{jsh1}) as a sum ${\bm j}_1 + {\bm j}_2 + {\bm j}_3$, where ${\bm j}_1$ is the contribution related with $\overline{\rho}^{(2)}_{n;0}$, and $ {\bm j}_2, {\bm j}_3$ are the contributions related with $\overline{\rho}^{(1)}_{n;0}$ and determined by the first and second terms in square brackets in Eq.~(\ref{ehk0}). One can check that, neglecting the Coulomb interaction, the currents $ {\bm j}_1$ and ${\bm j}_3$ cancel each other out. Therefore, with allowance for the electron-hole interaction, the sum $ {\bm j}_1 + {\bm j}_3$ is small, as compared to the current $ {\bm j}_2$, by the parameter $E_B/E_g$, and preserving this sum is an excess of accuracy, since while calculating the modified electron-hole states we neglected the terms that have such smallness. The allowance for these corrections in the calculation of the exciton states and the exciton oscillator strength has been made out in the work \cite{Leppenen} and is not carried out here.

Thus, the shift photocurrent, calculated taking into account the Coulomb interaction, is given by 
\begin{equation}
\label{shift_current_ee}
\bm{j}^{({\rm sh})}=e\dfrac{2\pi}{\hbar}  \sum_{n} {\bm R}_n |V_n|^2 \delta(E_n-\hbar\omega) \:,
\end{equation}
where $V_n = \langle s_e,s_h,{\bm k}| V | 0 \rangle$, ${\bm R}_n$ is an elementary charge shift induced by the optical transition
\begin{equation} \label{RRmu}
\bm{R}_n = - \dfrac{1}{|V_n|^2} \operatorname{Im}\left(V_n^*\sum\limits_{\bm{q}} C^{({\bm k})*}_{\bm q} \dfrac{\partial V_{c,s_e,{\bm q};v,-s_h,{\bm q}}}{\partial \bm{q}}\right)
\:.
\end{equation} 
Let us substitute Eqs.~(\ref{me_bulk}) and (\ref{MEs}) for the matrix elements into Eq.  (\ref{shift_current_ee}). Since the derivative of the matrix element (\ref{me_bulk}) is independent of ${\bm q}$, we can rewrite the sum over ${\bm q}$ as
\[
\sum\limits_{\bm{q}} C^{({\bm k})*}_{\bm q} \dfrac{\partial V_{c,s_e,{\bm q};v,-s_h,{\bm q}}}{\partial \bm{q}} = \sqrt{\cal{Z}} \frac{ \partial V_{ c, s_e, {\bm k}; v, - s_h ,{\bm k}}}{\partial {\bm k}  }\:.
\]
Finally, we obtain the second important result of the work
\begin{equation} \label{comp}
{\bm j}^{({\rm sh})}({\rm Coul}) = - e \frac{Q}{P} W =  {\cal Z} {\bm j}^{({\rm sh})}({\rm no\mbox{-}Coul})\:.
\end{equation} 
One can see that in the two-band model under consideration, the ratio of shift photocurrents calculated with and without allowance for the Coulomb interaction coincides with the similar ratio of light absorption coefficients.
\section{Comparison of the ballistic and shift contributions}
From Eq.~(\ref{jzf}) follows the frequency dependence of the ballistic photocurrent
\begin{equation} \label{llgg}
j^{({\rm bal})}_z = \frac{C}{a_B^2} \frac{k(\omega)}{1 - {\rm exp}[ - 2 \pi/k(\omega) a_B] } \:.
\end{equation}
where the wave vector $k(\omega)$ is defined in Eq.~(\ref{gE}) and the coefficient $C$ is independent of the effective mass $\mu$ and frequency $\omega$. The same frequency dependence has the expression for the current presented in Ref.~\cite{Entin1979}. However, in that formula, for the coinciding effective masses of the electron and hole, the mass $\mu$ is present only in the exponent, while the expression (\ref{llgg}) contains also the factor $\mu^2$ (due to $a_B^2$ in the denominator).

According to (\ref{jzf}) and (\ref{comp}), the ratio of the Coulomb ballistic and shift contributions to the current is described by 
\begin{equation} \label{balsh}
\frac{|j_z^{({\rm bal})}|}{|j_z^{({\rm sh})}|}=\frac23 \frac{\tau}{\hbar} \frac{\hbar^2 k}{\mu a_B} = \frac43 \frac{\tau}{\hbar} \sqrt{E_B (\hbar \omega - E_g)} = \frac23 \frac{\ell}{a_B}\:,
\end{equation}
where $\ell$ is the mean free path $\tau \hbar k /\mu$. Thus, we confirm the statement made in the paper \cite{SturmanUFN}: except for special cases of an extremely large value of the exciton Bohr radius (small effective mass, large permittivity) and a very short scattering time, the ballistic current dominates over the shift current. According to Eq.~(\ref{balsh}) the condition for this predominance is the inequality $\ell \gg a_B$ and not the inequality $\ell \gg a$ indicated in Ref.~\cite{SturmanUFN} ($a$ being the lattice constant). It should be stressed, however, that, in contrast to interband absorption, for intersubband transitions within one band, the shift and phonon ballistic mechanisms make decisive and comparable contributions to the LPGE \cite{Rasulov,Lyanda,Tarasenko}.

The opposite statement about the predominance of the shift contribution over the ballistic one is made in Refs.~\cite{Dai,RappeReview}. This may be due to the fact that the second term in Eq.~(8) in Ref.~\cite{Dai} or Eq.~(22) in Ref.~ \cite{RappeReview} includes an extra imaginary unit as a factor.

\section{Conclusion}
Within the framework of one band structure model of a bulk semiconductor, the ballistic and shift contributions to the linear photogalvanic effect, $j^{({\rm bal})}$ and $j^{({\rm sh})}$, respectively, are calculated. Both contributions are calculated taking into account the Coulomb electron-hole interaction. It is shown that in typical semiconductors the ballistic contribution significantly exceeds the shift contribution. The estimate for the ratio $j^{({\rm bal})}/j^{({\rm sh})}$ is given by $(\tau/\hbar)[E_B (\hbar \omega - E_g)]^{1/2}$. In the two-band model under consideration, the ratio of the shift current $j^{({\rm sh})}$ to the light absorption coefficient is independent of the frequency, whereas for the ballistic photocurrent this ratio increases monotonically with the increasing frequency according to the square root law $\sqrt{\hbar \omega - E_g}$ (with constant relaxation time $\tau$). We have considered a relatively simple two-band model, which allowed us to derive analytical formulas (\ref{jzf}) and (\ref{comp}). Using a more complex model would require numerical calculation of the Coulomb electron-hole functions.

In recent years, many publications have appeared, see Introduction, in which success has been achieved in the numerical calculation of the shift LPGE at interband transitions. An additional calculation 
of the ballistic photocurrent taking into account the Coulomb electron-hole interaction will allow to
obtain significantly larger values of the total photocurrent.

\acknowledgments

We acknowledge useful discussions with L.E. Golub, B.I. Sturman, S.A. Tarasenko and M.V. Entin.

The work is supported by the Russian Science Foundation grant N 22-12-00211.
\appendix

\section{Electric current of an electron-hole Coulomb pair} \label{Append1}
A pair of free electron and hole moving into opposite directions with velocities $\hbar {\bm q}/m^*$ and $(-\hbar {\bm q})/m^*$ carries a current
\[
{\bm j} = e \left[ \frac{\hbar {\bm q}}{m^*} - \left( - \frac{\hbar {\bm q}}{m^*}\right)\right] = e
\frac{\hbar {\bm q}}{\mu}.
\]
In the language of quantum physics, this means that the matrix element of the current operator between the states of free pairs is equal to
\begin{equation} \label{free0}
\langle s'_e, {\bm q}'; s'_h, - {\bm q}'; {\rm free} | \hat{\bm j} | s_e, {\bm q}; s_h, -{\bm q}; {\rm free} \rangle = e \frac{\hbar {\bm q}}{\mu} \delta_{{\bm q}{\bm q}'} \delta_{s_e s'_e} \delta_{s_h s'_h}\:.
\end{equation}

Using the expansion (\ref{exp2}) of the Coulomb wave function in terms of the states of non-interacting electron and hole, we calculate the matrix element between the two Coulomb states
\begin{eqnarray} \label{bmj}
&&\langle s'_e, s'_h, {\bm k}' | \hat{\bm j} | s_e, s_h, {\bm k} \rangle  
= \sum\limits_{{\bm q} {\bm q}'} C^{({\bm k}')*}_{{\bm q}'} C^{({\bm k})}_{\bm q} \langle s'_e, {\bm q}'; s'_h, - {\bm q}'; {\rm free} | \hat{\bm j} | s_e, {\bm q}; s_h, -{\bm q}; {\rm free} \rangle
\\ && \hspace{1.3 cm}= e \delta_{s_e s'_e} \delta_{s_h s'_h} \sum\limits_{{\bm q} {\bm q}'} C^{({\bm k}')*}_{{\bm q}'} C^{({\bm k})}_{\bm q} \frac{\hbar {\bm q}}{\mu}\delta_{{\bm q}{\bm q}'} =  e \delta_{s_e s'_e} \delta_{s_h s'_h} \sum\limits_{\bm q} C^{({\bm k}')*}_{\bm q}  \frac{\hbar {\bm q}}{\mu} C^{({\bm k})}_{\bm q}\nonumber \\ && \hspace{1.3 cm} = e \delta_{s_e s'_e} \delta_{s_h s'_h} \int \psi^{(+) *}_{{\bm k}'}({\bm r}) \left(- {\rm i} \frac{\hbar}{\mu} \frac{\partial}{\partial {\bm r}}\right) \psi^{(+) }_{\bm k}({\bm r}) d {\bm r} \:. \nonumber
\end{eqnarray}
Multiplying this matrix element by the density matrix and summing over the wave vectors and spin states, we obtain Eqs. (\ref{bal}), (\ref{k'vk}).

The matrix element $\langle 0 | \hat{\bm v} | s_e,s_h, {\bm k} \rangle$ in Eq. (\ref{sh}) is expressed through the Fourier components of $C^{({\bm k})}_{\bm q}$ as follows
\begin{equation} \label{free}
\langle 0 | \hat{\bm v} | s_e,s_h, {\bm k} \rangle = \sum\limits_{\bm q} C^{({\bm k})}_{\bm q} {\bm v}^*_{c,s_e; v,-s_h}({\bm q})\:,
\end{equation}
where the single-particle electron state $|v,-s_h, {\bm q} \rangle$ differs from the hole state $|h,s_h, -{\bm q} \rangle$ by the time inversion operation.
\section{Matrix element of coordinate between continuum states} \label{Append2}
Here we calculate the integral (\ref{r10}). First, we note that for radial functions of free motion
\[
R^{(0)}_{k0}(r) = 2 \frac{\sin{kr}}{r} \:,\: R^{(0)}_{k1}(r) = 2 \left( \frac{\sin{kr}}{kr^2} - \frac{\cos{kr}}{r} \right)
\]
the matrix elements of the coordinate $r$ have a singular form
\begin{eqnarray} \label{R0R0}
&&\int\limits_0^{\infty} R^0_{k1}(r) R^0_{k'0}(r) r^3 dr  =
2\pi \left[ \frac{\partial}{\partial k'} \delta(k' - k) + \frac{1}{k}  \delta(k' - k)\right]\:,  \\
&&\int\limits_0^{\infty} R^0_{k0}(r) R^0_{k'1}(r) r^3 dr  =
2\pi \left[-  \frac{\partial}{\partial k'} \delta(k' - k) + \frac{1}{k}  \delta(k' - k)\right]\:. \nonumber
\end{eqnarray}
Substitution of these expressions into (\ref{jzPQ}) instead of the integrals (\ref{r10}) does not lead to a photocurrent because of the identities
\[
(x' - x)^2 \frac{\partial \delta(x'-x)}{\partial x'}  =0\:,\: (x' - x)^2 \delta(x'-x) = 0\:.
\]
This is consistent with the statement that, without taking the Coulomb interaction into account, the linear ballistic current does not arise under direct interband transitions.

For the Coulomb functions $R_{kl}(r)$ the integral (\ref{r10}) is transformed to  \cite{Veniard,Komninos2012,Madajczyk}
\begin{equation} \label{Ik'k2}
I_{k1,k'0} = A \frac{\partial}{\partial k'} \delta(k' - k) + B  \delta(k' - k) + 2 \pi I^G_{k1,k'0} \:,
\end{equation}
where $A, B$ are functions of $k$, in particular, $A = 2 \pi[1 + (ka_B)^{-2} ]^{-1/2}$, see. e.g. 
Eq.~ (3.7) in Ref.~\cite{Veniard}, and the term $I^G_{k1,k'0}$ was calculated by Gordon in 1929 \cite{Gordon}. This term includes the factor $2 \pi$, since the integrand in Eq.  (\ref{r10}) contains radial functions in the Landau--Lifshitz normalization, which differ by a factor of $\sqrt{2 \pi}$ from the radial functions in Ref. \cite{Gordon}, see Eq. (\ref{relation}). As well as in the case of free pairs, the first two terms do not contribute to the photocurrent. Therefore, the ballistic photocurrent is determined by the Gordon integral, which we will represent in the following form
\begin{eqnarray} \label{zGordon2}
&& \hspace{3.5 cm}I^G_{k1,k'0} = \frac{f(k, k')}{(k' - k)^2} \:, \\
&& f(k, k') = {\rm i}
\frac{ 2 k k'}{(k + k')^2}  {\rm e}^{\frac{\pi}{2a_B}\left\vert \frac{1}{k'} - \frac{1}{k}\right\vert}\sqrt{ \frac{1 + \frac{1}{(k a_B)^2}}{k a_B\ \sinh{\frac{ \pi }{ k a_B} }\ k' a_B\ \sinh{\frac{ \pi }{ k' a_B} }}} \nonumber \\ && 
\left\{ \left\vert  \frac{ k - k' }{ k + k' } \right\vert^{- \frac{\rm i}{a_B} \left(\frac{1}{k}  - \frac{1}{k'} \right)  } \mbox{}_2F_1 \left( - \frac{\rm i}{k a_B} , 1 + \frac{\rm i}{k' a_B}, 2, \frac{4 kk'}{(k + k' )^2} \right)  \right.\nonumber \\ && \left.-  \left\vert \frac{k - k' }{ k + k' }\right\vert^{\frac{\rm i}{a_B} \left(\frac{1}{k}  - \frac{1}{k'} \right) } \mbox{}_2F_1\left(  \frac{\rm i}{k a_B} , 1 - \frac{\rm i}{k' a_B}, 2, \frac{4kk'}{(k + k' )^2} \right) \right\}\:, \nonumber
\end{eqnarray}
where $\mbox{}_2F_1(\alpha,\beta,\gamma;z)$ is the hypergeometric function. The integral $I^G_{k1,k'0}$ diverges as $k' \to k$. However, given the square of the energy difference in the numerator in the sum (\ref{jzPQ}), this sum converges, since the ratio
\begin{equation} \label{ratiomu}
\frac{ (E_{{\bm k}'} - E_{\bm k})^2}{(k'-k)^2}= \left( \frac{\hbar^2}{2 \mu} \right)^2 (k'+k)^2
\end{equation}
already has no singularity. The cancellation of squares $(k'-k)^2$ in the numerator and denominator allows us to perform a transformation with the appearance of an additional delta-function
\begin{equation} \label{delta}
\frac{1}{ (E_{{\bm k}'} - E_{\bm k})^2 + (\hbar/\tau)^2} = \frac{\pi \tau}{\hbar} \delta(E_{{\bm k}'} - E_{\bm k}) \:.
\end{equation}
Because of this delta-function, the variables $k, k'$ of the function $f(k,k')$ become equal, which significantly simplifies its form due to the identity
\begin{equation}
_2F_1 (\alpha, \beta, \gamma, 1) = \frac{\Gamma(\gamma) \Gamma(\gamma - \alpha - \beta)}{ \Gamma(\gamma - \alpha) \Gamma(\gamma - \beta)} \:,
\end{equation}
valid for $\textfrak{R} (\gamma) > \textfrak{R}(\alpha + \beta)$. For the difference of hypergeometric functions we obtain
\begin{eqnarray}
&& \mbox{}_2F_1 \left( - \frac{\rm i}{k a_B} , 1 + \frac{\rm i}{k' a_B}, 2, 1 \right) - \mbox{}_2F_1 \left(  \frac{\rm i}{k a_B} , 1 - \frac{\rm i}{k' a_B}, 2, 1 \right) \\ &&= - \frac{2 {\rm i}}{k a_B}  \frac{1}{1 + \frac{1}{k^2a^2_B} } \frac{1}{\left| \Gamma( 1 + \frac{\rm i}{ka_b})\right|^2} = - \frac{2 {\rm i}}{\pi}  \frac{\sinh{(\pi/ka_B)}}{1 + (ka_B)^{-2} }\:, \nonumber
\end{eqnarray}
because
\[
\left| \Gamma \left( 1 + \frac{\rm i}{ka_b} \right)\right|^2 = \frac{\pi/k a_B }{\sinh{(\pi/ka_B)}}\:.
\]
As a result we have
\begin{equation} \label{fkk}
f(k, k) = \frac{1}{  \pi} \frac{1}{ \sqrt{ (k a_B)^2 + 1 }}  \:.
\end{equation}
It is evident that the value $f(k,k)$ differs from zero only if the Coulomb interaction is taken into account; for a free electron-hole pair, for which $a_B \to \infty$, this value tends to zero and the ballistic photocurrent vanishes.

\end{document}